\documentclass[twocolumn,secnumarabic,amssymb, nobibnotes, aps, prd,superscriptaddress]{revtex4-2}
\usepackage{graphicx}
\usepackage{color}
\usepackage{amsmath}
\usepackage{gensymb}
\usepackage[left,modulo]{lineno}
\usepackage{upgreek}

\usepackage{siunitx}

\newcommand{\REV}[1]{#1}

\begin{document}
\title{Onset of glacier tables}

\author{Marceau H\'enot}
\affiliation{Univ Lyon, ENS de Lyon, Univ Claude Bernard, CNRS, Laboratoire de Physique, F-69342 Lyon, France}

\author{Nicolas Plihon}
\affiliation{Univ Lyon, ENS de Lyon, Univ Claude Bernard, CNRS, Laboratoire de Physique, F-69342 Lyon, France}

\author{Nicolas Taberlet}
\email[Corresponding author: ]{nicolas.taberlet@ens-lyon.fr}
\affiliation{Univ Lyon, ENS de Lyon, Univ Claude Bernard, CNRS, Laboratoire de Physique, F-69342 Lyon, France}

\date{\today}
\begin{abstract}
 
  A glacier table consists of a rock supported by a slender column of ice and form naturally on glaciers. We investigate the onset of their formation at a smaller scale in a controlled environment. Depending on the size and thermal conductivity of a cap, it can either form of a table standing on an ice foot, or sink into the ice block. A one-dimension conduction model shows that the differential ice melting is controlled by a competition between two effects: a geometrical amplification, and a heat flux reduction due to the higher temperature of the cap as compared to the ice. Our model captures the transition between the two regimes and identifies a dimensionless number which controls the onset of glacier tables formation.
 
\end{abstract}
\maketitle

%%%%%%%%%%%%%%%%%%%%%%%%%%%%%%%%%%%%%%%%
%%%%%%%%%%%%%%%%%%%%%%%%%%%%%%%%%%%%%%%%
\textbf{Introduction.} 

Differential ablation in Nature is an efficient drive for the formation of spectacular structures or patterns. For instance, \REV{the formation of meandering rivers or surface patterns known as rillenkarren can result from dissolution of soluble rocks~\cite{cohen2016erosion,cohen2020buoyancy,guerin2020streamwise}}. Furthermore, fairy chimneys appear when a stone protects a narrow column of sedimentary rock from erosion~\cite{Bruthans2014, Turkington2005}. While mechanical ablation is indeed a widely spread mechanism, differential physical phase change may also be an efficient differential ablation process. This is for instance the case in sublimation-driven pattern formation, ranging from penitentes on Earth due to ice sublimation~\cite{Mangold2011,Bergeron2006,Claudin2015} to landscape formation on Mars due to carbon dioxide sublimation~\cite{Head2003,Masse2010,McKeown2017} or on Pluto~\cite{Moore2017}. Differential melting of ice leads to the formation of glacier tables (see Fig.~\ref{photos}.a), which are structures frequently encountered on Alpine glaciers. They typically consist of a meter-size rock supported by a column of ice~\cite{Agassiz1840} and can last several years~\cite{Bouillette1934}. Glacier tables form because the ablation rate of the ice is lower under the rock than at the air-ice interface. However, under other circumstances, a stone placed on the ice surface may instead sink in~\cite{McIntyre1984}. Dirt layers on a glacier and differential melting can also lead to the formation of dirt cones~\cite{Campbell1975,drewry_1972} for which shortwave reflection is important~\cite{betterton_theory_2001}, as for suncups~\cite{rhodes_armstrong_warren_1987, mitchell_growth_2010}. Understanding the dynamics of differential melting on glacier require the modeling of ice melting under various conditions (including under a debris layer) and of the energy balance on a glacier. While the melting of ice submerged in water~\cite{Sugawara1975, Yen1980,FUKUSAKO199390} showed the importance of convection and instabilities arising from the density inversion of water, ice melting in air has drawn far less attention. A shielding effect from the water film covering an ice block was ruled out  theoretically~\cite{Roberts1958} and the effect of the heat flux from condensation of humid air was recently investigated~\cite{Chandramohan2016}.
%On the other hand, the energy balance responsible for the ice melting on a glacier is influenced by various physical processes such as wind, rain, shortwave and longwave radiation~\cite{hock_glacier_2005} with an important time variability which gives rise to complex phenomena.

%%%%%%%%%%%%%%%%%%%%%%%%%%%%%%%%%%%%%%%%
\begin{figure}[htbp]
  \centering
  \includegraphics[width=8.6cm]{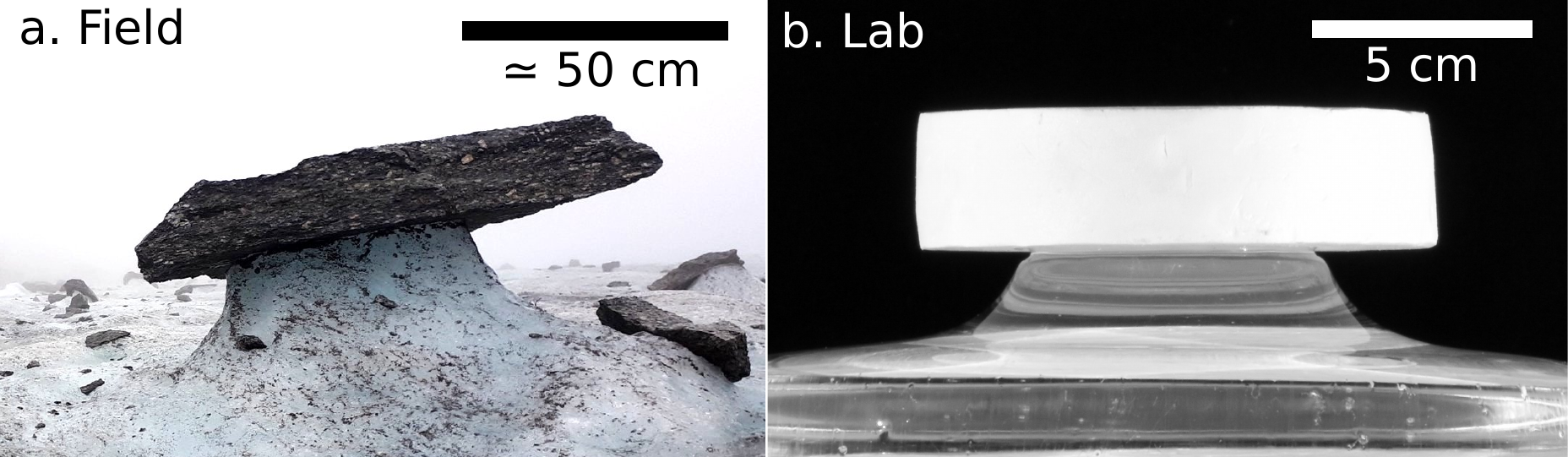}
 \caption{(a) Picture of a glacier table taken in June 2019 on the Mer de Glace in the French Alps. The table diameter is about 1~m.  (b) Artificial glacier table made of extruded polystyrene (XPS), 10~h after being put on a flat ice surface.}
  \label{photos}
\end{figure}
%%%%%%%%%%%%%%%%%%%%%%%%%%%%%%%%%%%%%%%%

The purpose of this Letter is to determine, from a physical point of view, the onset of formation of glacier tables under a controlled environment. More specifically, we address under which conditions (thermal properties of the material, dimensions and aspect ratio) a cylindrical object placed on an ice surface will initially either sink in or seemingly rise. In order to identify the parameters controlling the ice melting process in air, we first studied the melting of inclined ice plates, in a controlled environment (constant temperature, humidity, absence of wind), \REV{and we show that ice melting due to convection is the leading factor.}
We then investigated the behavior of cylindrical caps on a flat ice surface. These experimental results are well fitted using an analytical thermal conduction model, from which a single dimensionless number is shown to control the onset of glacier table formation. Conditions for the formation of natural glacier tables are finally extrapolated from laboratory-scale experiments.

%%%%%%%%%%%%%%%%%%%%%%%%%%%%%%%%%%%%%%%%
%%%%%%%%%%%%%%%%%%%%%%%%%%%%%%%%%%%%%%%%

\textbf{Ice melting in air.} The melting of inclined ice plates in air was studied experimentally under controlled conditions. A slab of clear ice (\REV{see suppl. mat.}) is laid onto an inclined sheet of Styrofoam, which entirely prevents the ice from melting from underneath. 
The height of the ice slab $e(x)$ was monitored from series of photographs, taken every 8 min and was extracted using an image-processing Python package (see details in suppl. mat.). The local heat flux $q_\mathrm{total}(x)$ as a function of the distance $x$ from the leading edge of the plate (the highest point from which cold air will flow down the slope), was deduced from the measurement of the local melting rate $v_\mathrm{ice}(x)=\mathrm{d}e(x)/\mathrm{d}t =q(x)/\mathcal{L}_\mathrm{fus}$ where $\mathcal{L}_\mathrm{fus} = 303$~MJ$\cdot$m$^{-3}$ is the enthalpy of fusion of ice. The heat flux profiles are shown in Fig.~\ref{fonte_plan_incline}.a for different inclination angles $\theta$ ranging from \ang{11} to \ang{85}. The heat flux is clearly affected by the inclination and strongly increases near the leading edge.

%%%%%%%%%%%%%%%%%%%%%%%%%%%%%%%%%%%%%%%%
\begin{figure}[htbp]
  \centering
  \includegraphics[width=8.6cm]{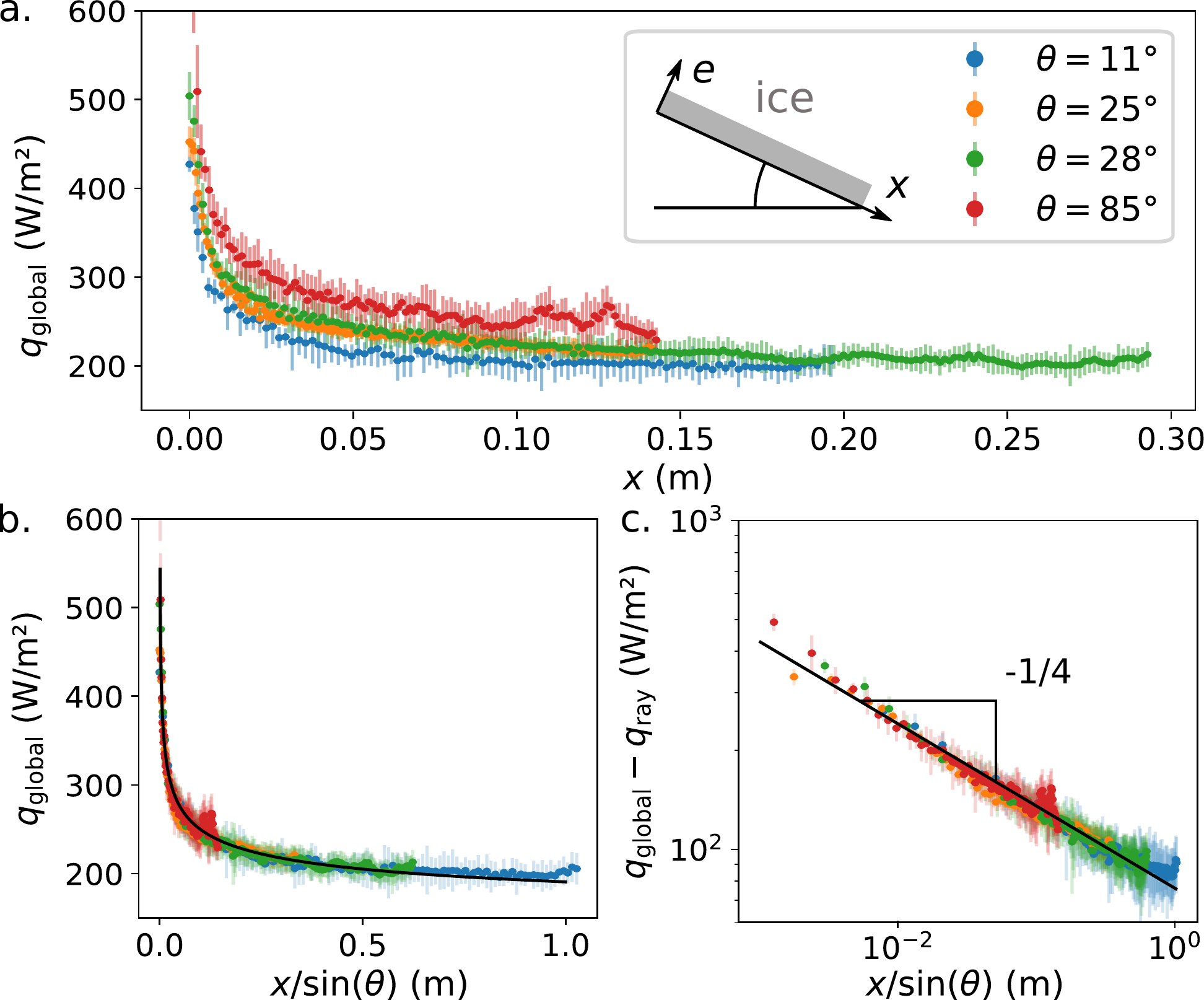}
 \caption{ Heat flux $q_\mathrm{total}$ received by an ice plate inclined with an angle $\theta$ with respect to the horizontal (a) as a function of the distance $x$ from the leading edge and (b) as a function of $x/\sin \theta$. \REV{Graph (c)} shows $q-q_\mathrm{ray}$ as a function of $x/\sin \theta$. \REV{The black lines correspond to the expressions of $q_\mathrm{total}$ (b) and $q_\mathrm{conv}$ (c) given in eq.~\ref{eq_conv} with an adjustable numerical prefactor $A=0.61$.}}
  \label{fonte_plan_incline}
\end{figure}
%%%%%%%%%%%%%%%%%%%%%%%%%%%%%%%%%%%%%%%%

As the ice melts, its surface is covered by a thin film of liquid water. Assuming that this film originates only from the melting of ice and not by condensation, the volumetric flow rate (per unit of width) can be computed from the mean melting rate: $\REV{\langle \dot{V}_\mathrm{w} \rangle = \frac{1}{L} \int_0^L\langle v_\mathrm{ice}\rangle x \mathrm{d}x =} \langle v_\mathrm{ice}\rangle L/2 \approx 10^{-7}$~m$^2\cdot$s$^{-1}$ where $L$ is the length of the ice plate. 
\REV{Note here that the flow rate $\langle \dot{V}_\mathrm{w}\rangle$ and the water velocity vary with $x$. The typical thickness of the water layer $e_w$ is estimated from the flow rate $\langle \dot{V}_\mathrm{w} \rangle$ assuming a free surface flow with a semi-parabolic velocity profile~\cite{Batchelor} as $e_\mathrm{w}= \sqrt[3]{3\nu_\mathrm{w}\langle \dot{V}_\mathrm{w}\rangle  / g } \approx 40~\upmu$m, where $g$ is the gravitational acceleration and $\nu_\mathrm{w}$ 
is the kinematic viscosity of water. This thickness is in agreement with previous measurements~\cite{Chandramohan2016}. The \REV{maximum} typical water velocity $v_\mathrm{w}$ at the center of the plate ($x=L/2$) is obtained for a vertical plate  ($\theta = 90$\degree) as $v_\mathrm{w} = g e_\mathrm{w}^2/(2\nu_\mathrm{w}) \approx 4$~mm$\cdot$s$^{-1}$. On the edges (top and sides) we expect the water flow rate and thus the thickness and velocity to vanish.}
The typical time of thermal diffusion through the water film $\tau_\mathrm{diff} =  e_\mathrm{w}^2/D_\mathrm{w} \approx 0.01$~s is much smaller than the advection time $\tau_\mathrm{adv} = L/(2v_\mathrm{w}) \approx 30$~s where $D_\mathrm{w}$ is the thermal diffusivity of water. 
Therefore the liquid film is not expected to play any role in the energy balance (thermal conduction through the thin film of liquid water is quick) and the incoming heat flux at the ice surface is equal to the local heat flux at the top of the liquid layer (from the ambient air), consistently with previous analysis~\cite{Roberts1958}. 
 Note that the heat flux is two orders of magnitude higher than numerical predictions assuming a leading effect of condensation for a vertical ice plate~\cite{Chandramohan2016}, thus ruling out condensation as the main mechanism governing the melting.

The heat sources in the experiment are infrared radiations coming from the walls at room temperature \REV{($q_\mathrm{ray}$)} and natural air convection \REV{($q_\mathrm{conv}$)}. \REV{The total heat flux is denoted as $q_\mathrm{total} = q_\mathrm{ray} + q_\mathrm{conv}$.}
The ice is opaque to far infrared radiations which are absorbed in the first tens of micrometers near the surface~\cite{warren_optical_2008}. For such wavelengths the emissivity of ice and of the chamber walls can be approximated to 1~\cite{Emissivity}. The net radiative flux received by the ice is then $q_\mathrm{ray} = \sigma (T_\mathrm{room}^4 - T_\mathrm{ice}^4)$ where $\sigma$ is the Stefan–Boltzmann constant and $T_\mathrm{room} = 295$~K is the room temperature. This gives $q_\mathrm{ray} = 110 $~W$\cdot$m$^{-2}$ which constitutes, far from the leading edge of the plate, roughly half of the measured heat flux.
The convective heat flux may be estimated from a numerous studies of natural convective heat transfer between air and a plate with a homogeneous temperature~\cite{Rich1953,Hassan1970, FUJII1972, CHURCHILL1975}.
A theoretical assumption~\cite{Rich1953} states that the laminar convective flow of air on an inclined plate is equivalent to the problem of a vertical plate where the $g$ is replaced by its projection on the plate surface $g\sin \theta$, which was later experimentally verified~\cite{Hassan1970, FUJII1972, CHURCHILL1975}. The analytical calculation of the air boundary layer profile gives a power law dependence of the convective flux with the position~\cite{lienhard_heat_2019} : 
\begin{equation}
     q_\mathrm{conv}(x) = A \lambda_\mathrm{air}(T_\mathrm{room}-T_\mathrm{ice})^{5/4} \left( \frac{g \beta_\mathrm{air}}{\nu_\mathrm{air} D_\mathrm{air}} \frac{\sin{\theta}}{x}\right)^{1/4}
    \label{eq_conv}
\end{equation}
\noindent where \REV{$A$ is a numerical prefactor}, $\lambda_\mathrm{air} = 0.025$~W$\cdot$m$^{-1}\cdot$K$^{-1}$, $D_\mathrm{air} = 2.0 \cdot 10^{-5}$~m$\cdot$s$^{-2}$, $\beta_\mathrm{air} = 1/T_\mathrm{air}$ and $\nu_\mathrm{air} = 1.4 \cdot 10^{-5}$~m$\cdot$s$^{-2}$ are the thermal conductivity, thermal diffusivity, thermal expansion coefficient and kinematic viscosity of air~\cite{handbook_thermal}.

Figure~\ref{fonte_plan_incline}.b shows the evolution of the heat flux as a function of $x/\sin\theta$, and, as expected, all data collapse on one master curve. The black solid line corresponds to $q_\mathrm{ray} + q_\mathrm{conv}(x/\sin\theta)$ with the room temperature measured independently ($T_\mathrm{room}=$~\SI{21.7}{\celsius}), all thermal parameters being extracted from the literature~\cite{handbook_thermal} \REV{and an adjusted numerical prefactor $A=0.61$ very close to the theoretical (0.41)~\cite{lienhard_heat_2019} and experimental (0.35-0.56)~\cite{Hassan1970, FUJII1972, CHURCHILL1975} values reported in the literature (see suppl. mat. and Refs. [40-45] therein)}. The convective heat flux, estimated as  $q(x/\sin \theta) - q_\mathrm{ray}$, is displayed as a function of  $x/\sin\theta$ in Fig.~\ref{fonte_plan_incline}.c and the very good agreement between the experimental results and the model demonstrates that the melting of an ice sheet in air is controlled both by the infrared radiations coming from the room and by the natural convection of the air. In particular, the smoothing of edges (or corners) that can be observed during the melt of an ice block in air can be attributed to to the divergence of the heat flux where the air thermal boundary layer vanishes.

%%%%%%%%%%%%%%%%%%%%%%%%%%%%%%%%%%%%%%%%
%%%%%%%%%%%%%%%%%%%%%%%%%%%%%%%%%%%%%%%%

\textbf{Artificial glacier tables.} We now investigate the formation of artificial glacier tables in a similar controlled environment. As can be seen in Fig.~\ref{photos}.b, a small-scale artificial glacier table can be produced by laying a cylindrical extruded polystyrene (XPS) cap on an ice surface for a few hours. In the following $H$ denotes the height of the cylinder and $R$ its radius. The aspect ratio of the cap is defined as $\beta = H/(2R)$. Experiments using cylinders made of five materials are reported in this Letter: extruded polystyrene (XPS), rigid polyvinyl chloride (PVC), plaster, cement and granite, with thermal conductivity $\lambda$ varying over two orders of magnitude (see suppl. mat.). 

%%%%%%%%%%%%%%%%%%%%%%%%%%%%%%%%%%%%%%%%
\begin{figure}[htbp]
  \centering
  \includegraphics[width=8.6cm]{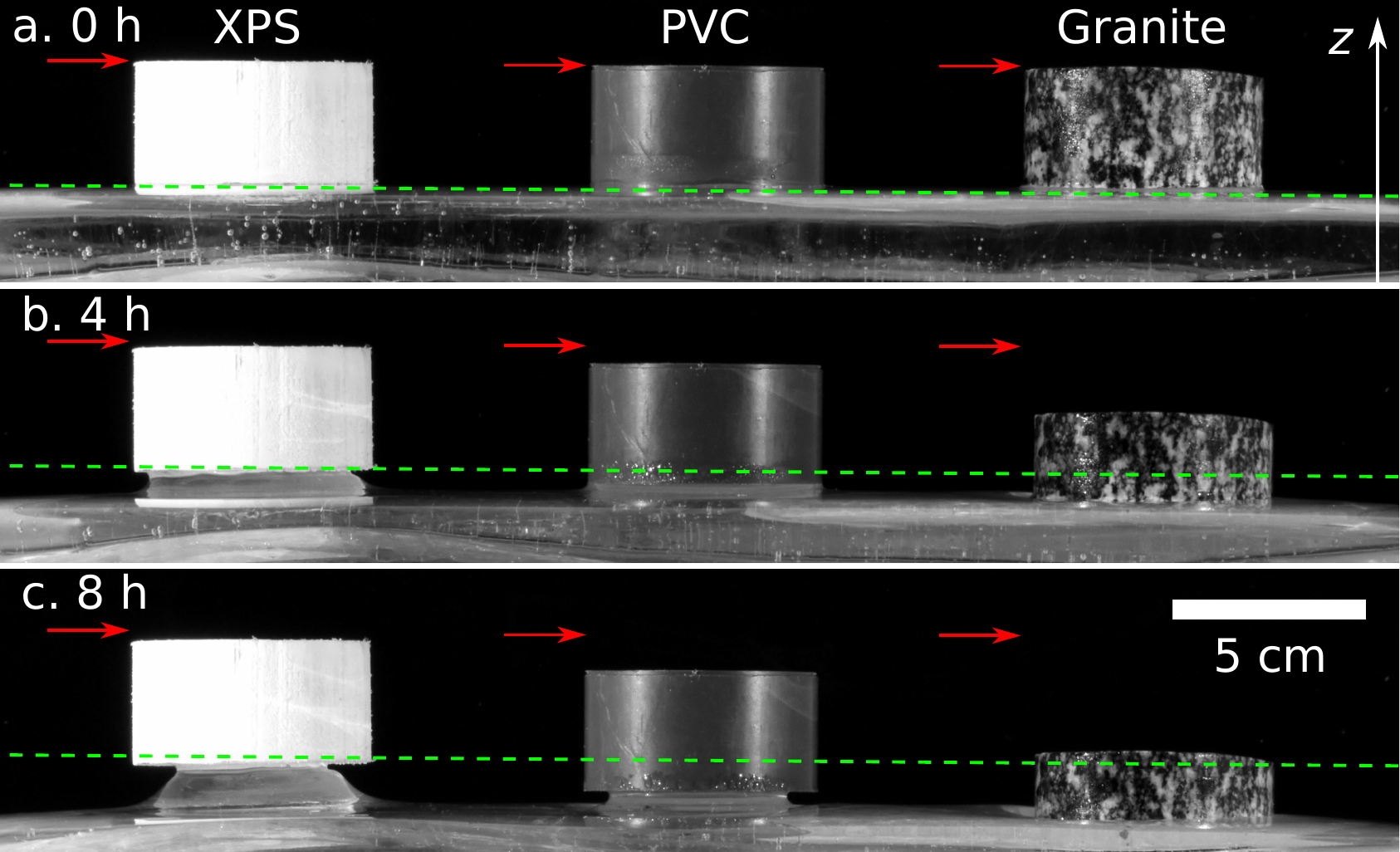}
 \caption{Cylindrical caps with aspect ratio $\beta=1/2$ made of XPS, PVC and granite, in the initial state (a), after 4~h (b) and 8~h (c). The red arrows denote the initial vertical positions of the top of the caps. The green dashed line corresponds to the initial position of the ice surface.}
  \label{3caps}
\end{figure}
%%%%%%%%%%%%%%%%%%%%%%%%%%%%%%%%%%%%%%%%

Figure~\ref{3caps} illustrates the evolution of a melting ice sheet on which caps are initially laid. Depending on the thermal properties of the material used, the outcome greatly varies. Indeed, under the XPS cap, the ice ablation rate is almost zero and a table forms. The melting under the PVC cap is clearly visible but the ablation rate stays lower under the cap than at the air interface, thus leading to the formation of a less protruding table. In the case of the granite however, the ablation rate is higher under the cap than at the air interface and the stone sinks into the ice sheet. When the cap forms a table, the ice column underneath it becomes thinner and thinner over time, and the cap eventually falls off. In the present Letter, the focus is on the onset of the formation of the table (the first 2~h), when the bottom of the cap is still fully in contact with the ice, and the morphogenesis of the foot over its lifetime is left for future work.
Figure~\ref{plot_beta1sur2} shows the evolution of the vertical position of the top of caps (for three of the tested materials) along with the position of the ice surface far from the cap. These evolutions are linear meaning that the system quickly reaches a steady state (within minutes) and that the ice ablation rate under the cap (and far from it) is constant over time. The vertical velocity of the cap $v_\mathrm{cap}$, and the velocity of the ice at the air interface $v_\mathrm{ice}$ (far from the cap), are measured from these evolutions, during the same experiment. The ability for a cap to form a table in early stages is estimated from the ratio $v_\mathrm{cap}/v_\mathrm{ice}$: if this ratio exceeds unity, the cap sinks into the ice block, whereas when it is less than unity, a table will initially form. The lower the ratio, the quicker the formation occurs.

%%%%%%%%%%%%%%%%%%%%%%%%%%%%%%%%%%%%%%%%
\begin{figure}[htbp]
  \centering
  \includegraphics[width=8.6cm]{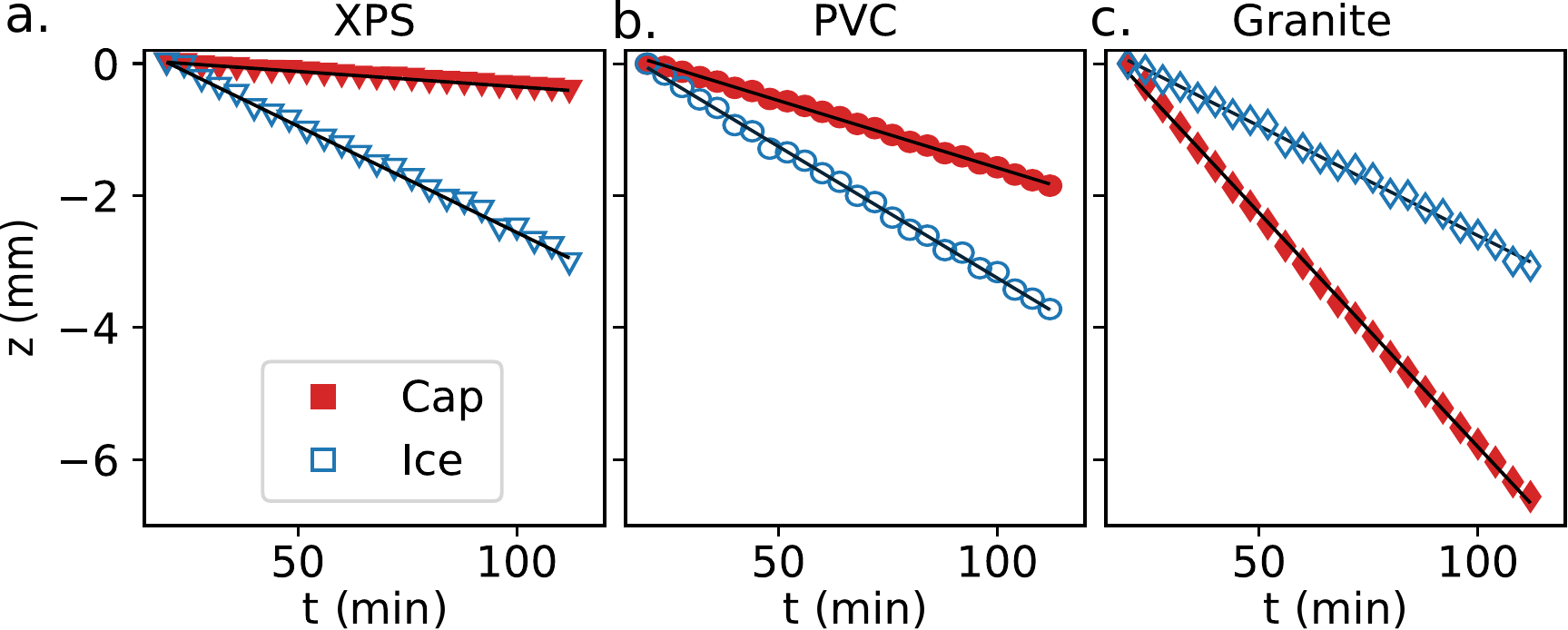}
 \caption{Vertical position of caps (red solid markers) and of the air-ice interface (blue empty markers, see suppl. mat.) as a function of time for caps made of a) XPS, b), PVC, and c) granite with an aspect ratio $\beta=1/2$ and radius $R = 42$~mm, 41~mm and 31~mm respectively. }
  \label{plot_beta1sur2}
\end{figure}
%%%%%%%%%%%%%%%%%%%%%%%%%%%%%%%%%%%%%%%%

The fact that caps of similar sizes but various thermal properties can display such opposite behaviors (forming a table or sinking) shows that there exist two competing effects governing the ice melting under the cap. On the one hand, there is a geometrical amplification effect: the heat flux coming from the environment to the cap, that eventually goes to the ice, is received on a higher surface $S_\mathrm{air-cap} = \pi R^2 + 2\pi R H$ than the contact area between the ice and the cap $S_\mathrm{ice-cap} = \pi R^2$. The geometrical amplification coefficient expresses $S_\mathrm{air-cap}/S_\mathrm{ice-cap} = 1+4\beta$. On the other hand, there is an attenuation effect of the heat flux because the temperature $T_\mathrm{cap}$ of the cap-air interface is higher than $T_\mathrm{ice}$. This reduces the radiative and the convective heat flux received by the cap compared to the one received by the ice as they both depend on the difference between the surface temperature and $T_\mathrm{room}$. In the following we introduce a one-dimensional model which account for both effects. 

%Let  $q = q_\mathrm{ray} + q_\mathrm{conv}$ be the heat flux received by the cap from the environment.
%A linear expansion of the total heat flux received by the cap from the environment gives $q_\mathrm{total}(T_\mathrm{cap}) = q_\mathrm{ray}(T_\mathrm{cap}) + q_\mathrm{conv}(T_\mathrm{cap}) = (4\sigma T_\mathrm{room}^3+h_\mathrm{conv})(T_\mathrm{room} - T_\mathrm{cap})$, \REV{where the weak temperature dependence of $h_\mathrm{conv}$ is neglected}.  An \REV{effective} heat transfer coefficient $h_\mathrm{eff} = 4\sigma T_\mathrm{room}^3+h_\mathrm{conv}$ can be defined taking into account both mechanisms. 
\REV{An effective heat transfer coefficient $h_\mathrm{eff}$ is introduced from a linear expansion of the total heat flux received by the cap from the environment $q_\mathrm{total}(T_\mathrm{cap}) = q_\mathrm{ray}(T_\mathrm{cap}) + q_\mathrm{conv}(T_\mathrm{cap}) = h_\mathrm{eff}(T_\mathrm{room} - T_\mathrm{cap})$, with $h_\mathrm{eff} = 4\sigma T_\mathrm{room}^3+h_\mathrm{conv}$ when the weak temperature dependence of $h_\mathrm{conv}$ is neglected.}
This coefficient, which is independent of the thermal properties of the materials, is therefore the same for the ice and for all caps. The measurements of the ice melting velocity in Fig.~\ref{plot_beta1sur2} lead to $h_\mathrm{eff} = 9.1\pm 0.1$~W$\cdot$K$^{-1}\cdot$m$^{-2}$, which is consistent with results found in the literature~\cite{lienhard_heat_2019}.

The heat fluxes received respectively by the cap and by the ice far away from the cap are labeled $q_\mathrm{air\rightarrow cap} = h_\mathrm{eff}(T_\mathrm{room}-T_\mathrm{cap})$ and $q_\mathrm{air\rightarrow ice} = h_\mathrm{eff}(T_\mathrm{room}-T_\mathrm{ice})$. The ice under the cap receives a heat flux $q_\mathrm{cap\rightarrow ice} = (1+4\beta) q_\mathrm{air\rightarrow cap}$ due to the geometrical amplification. Moreover, the heat flux going through the cap is computed using a 1D steady-state conduction model: $q_\mathrm{cap\rightarrow ice} = \lambda\frac{T_\mathrm{cap}-T_\mathrm{ice}}{d}$ where $d$ is a typical length scale. For an infinitely wide cap (i.e. $\beta =0$), the characteristic length $d$ would directly be given by the height of the cap $H$.  Here, the 3D geometry of the cap is taken into account by assuming that $d$ is proportional to the ratio between the volume of the cap and the surface in contact with air : $d = \eta V_\mathrm{cap}/S_\mathrm{air-cap} = \eta\ 2R\beta/(1+4\beta)$ where $\eta$ is a numerical prefactor. This allows one to express temperature difference between the cap and the ice $T_\mathrm{cap}-T_\mathrm{ice}=(T_\mathrm{room}-T_\mathrm{ice})/(1+\frac{\lambda}{2\eta \beta h_\mathrm{eff} R})$ and to estimate the ratio between the heat fluxes responsible for the ice melting under the cap and far from the cap. This ratio is also the ratio of the melting velocities under the cap and far from the cap:
\begin{equation}
     \frac{v_\mathrm{cap}}{v_\mathrm{ice}}=\frac{q_\mathrm{cap\rightarrow ice}}{q_\mathrm{air\rightarrow ice}} = \frac{1+4\beta}{1+2\eta\beta\frac{h_\mathrm{eff}R }{\lambda}}
     \label{model1}
\end{equation}
Note that the dimensionless number $\frac{h_\mathrm{eff}R}{\lambda}$ which appears in eq. (2) is the only parameter accounting for the thermal properties of the material. Moreover, as the system quickly reaches a steady state, the only relevant thermal property of a given material is its thermal conductivity, while its heat capacity only plays a role in a rapid transient regime.

%%%%%%%%%%%%%%%%%%%%%%%%%%%%%%%%%%%%%%%%
\begin{figure}[htbp]
  \centering
  \includegraphics[width=8.6cm]{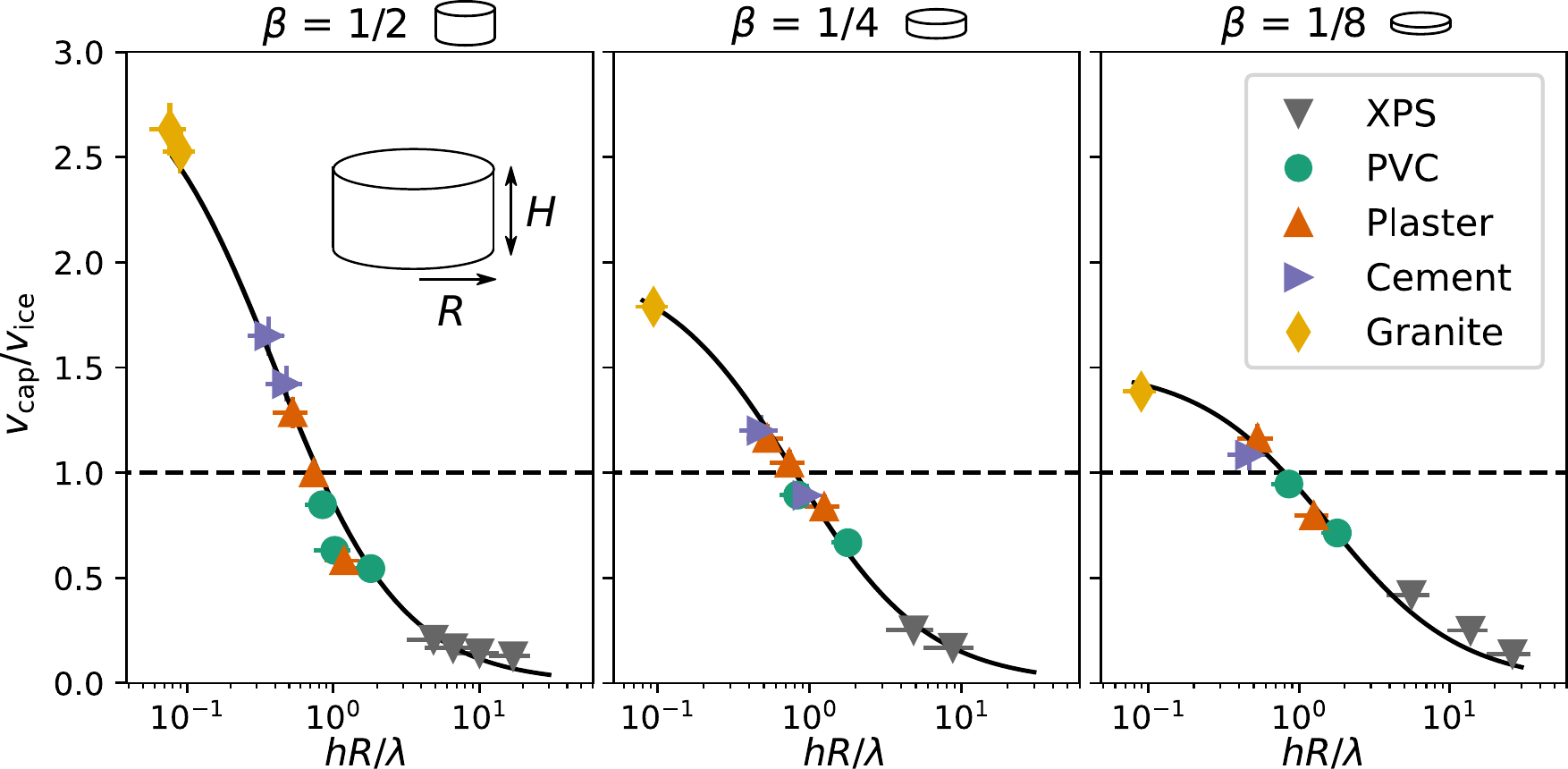}
 \caption{Ratio $v_\mathrm{cap}/v_\mathrm{ice}$ as a function of the dimensionless number $h_\mathrm{eff}R/\lambda$ for three values of the aspect ratio $\beta=H/(2R)$. The solid black lines correspond to Eq.~\ref{model1} with $\eta = 2.5$.}
  \label{bilan}
\end{figure}
%%%%%%%%%%%%%%%%%%%%%%%%%%%%%%%%%%%%%%%%

%%%%%%%%%%%%%%%
%%%%%%%%%%%%%%%
%%%%%%%%%%%%%%%
%%%%%%%%%%%%%%%
%%%%%%%%%%%%%%%

The velocities ratio $v_\mathrm{cap}/v_\mathrm{ice}$ for cylindrical caps made of XPS, PVC, plaster, cement and granite are shown as a function of $\frac{h_\mathrm{eff}R}{\lambda}$ in Fig.~\ref{bilan}, for three values of the aspect ratio $\beta$. The dimensionless number $\frac{h_\mathrm{eff}R}{\lambda}$ varies over three decades by changing the radius $R$ of the cylinder and its thermal conductivities $\lambda$ - which were measured in the case of PVC, plaster,cement and granite and taken from the literature for XPS (see suppl. mat.).
For each aspect ratio $\beta$, the two regimes (sinking and table formation) are visible. Note that the transition ($v_\mathrm{cap} = v_\mathrm{ice}$) is observed to occur experimentally for $\frac{h_\mathrm{eff}R}{\lambda} = 0.8$, independently of the value of $\beta$. In Eq.~\ref{model1}, the transition occurs for $\frac{\eta}{2}\frac{h_\mathrm{eff}R}{\lambda} = 1$ and our experimental observations lead to $\eta=2.5$. The complete prediction of  Eq.~\ref{model1} is shown as black lines in Fig. 5, in excellent agreement with the experimental data, which validates the simple assumptions of our model. 

From an extrapolation of our modeling, it is possible to estimate the critical size of rocks from which a glacier table can form on a glacier.
 The complexity of the energy balance on a glacier can be tackled using a temperature index model~\cite{hock_glacier_2005}, which relies on the observation that the net heat flux is strongly correlated to the temperature of the air above the glacier. Assuming a linear relationship leads to $v_\mathrm{ice} = \mathrm{DDF} (T_\mathrm{air}-T_\mathrm{ice})$. The average degree day factor $\mathrm{DDF}$ is reported to vary roughly between 5 and 10~mm$\cdot$day$^{-1}\cdot$K$^{-1}$~\cite{hock_temperature_2003} which corresponds to an effective heat exchange coefficient  $h_\mathrm{eff}~\REV{=\mathrm{DDF}\mathcal{L}_\mathrm{fus}/(24 \times 3600)}=20-40$~W$\cdot$K$^{-1}\cdot$m$^{-2}$. For granite rocks, this leads to a critical diameter of $D_\mathrm{lim}=1.6\lambda/h_\mathrm{eff}= 10-20$~cm which is consistent with the observation that the glacier tables usually have a metric size. 

%%%%%%%%%%%%%%%%%%%%%%%%%%%%%%%%%%%%%%%%
%%%%%%%%%%%%%%%%%%%%%%%%%%%%%%%%%%%%%%%%

\textbf{Conclusion and perspectives.}
We were able to produce small scale artificial glacier tables in a controlled environment in which the mechanisms of the heat transfer were identified. By comparing the experimental results with an analytical heat conduction model, we showed that the table formation in these conditions is controlled by a single dimensionless number, $\frac{h_\mathrm{eff}R}{\lambda}$ which takes into account the properties of the material laid on the ice. In the case of a small cap with high thermal conductivity, the geometrical amplification of the heat flux causes the cap to sink into the ice. In contrast, a large cap with low thermal conductivity will form a table as the higher temperature of the cap reduces the net heat flux to the ice below. From this, the critical rock size delimiting these two regimes on a glacier can be estimated to be of the order of tens of centimeters which is consistent with field observations. The question remains on how the direct or indirect solar irradiation, the wind, the diurnal variability, or the porosity of the ice and debris layer can affect this process in natural conditions~\cite{hock_glacier_2005, evatt_glacial_2015, reid_energy-balance_2010, ostrem1959}. Our approach should also be extended to the study of the overall shape of glacier tables throughout their lifetime. \REV{with a more complete understanding, glacier table might be used as proxies to estimate environmental parameters such as effective heat transfer coefficient or glacier ablation.}

%%%%%%%%%%%%%%%%%%%%%%%%%%%%%%%%%%%%%%%%
%%%%%%%%%%%%%%%%%%%%%%%%%%%%%%%%%%%%%%%% 
\textbf{Acknowledgements.}
The authors are grateful to Marine Vicet, J\'er\'emy Vessaire and Vincent Langlois for fruitful discussions, to the F\'ed\'eration de Recherche  Marie Andr\'e Amp\`ere and to the Laboratoire de Physique at the ENS de Lyon for financial support, and to Gabriel B. Plihon for providing geological samples.

%\bibliography{bibliographie}

%

\newpage

\onecolumngrid

\begin{center}
\textbf{\Large Supplementary materials}
\end{center}

\author{Marceau H\'enot}
\affiliation{Univ Lyon, Ens de Lyon, Univ Claude Bernard, CNRS, Laboratoire de Physique, F-69342 Lyon, France}

\author{Nicolas Plihon}
\affiliation{Univ Lyon, Ens de Lyon, Univ Claude Bernard, CNRS, Laboratoire de Physique, F-69342 Lyon, France}

\author{Nicolas Taberlet}
\affiliation{Univ Lyon, Ens de Lyon, Univ Claude Bernard, CNRS, Laboratoire de Physique, F-69342 Lyon, France}

\maketitle
%%%%%%%%%%%%%%%%%%%%%%%%%%%%%%%%%%%%%%%%
%%%%%%%%%%%%%%%%%%%%%%%%%%%%%%%%%%%%%%%%

%%%%%%%%%%%%%%%%%%%%%%%%%%%%%%%%%%%%%%%%
%%%%%%%%%%%%%%%%%%%%%%%%%%%%%%%%%%%%%%%%
\textbf{Materials and methods} 

Clear ice blocks were made using a cylindrical plastic container isolated at the bottom and on the sides by a 5-10~cm thick wool layer, and were filled with tap water and put in a \SI{-30}{\celsius} domestic freezer. The freezing front progress from the top free surface of the water towards the bottom. The container is removed from the freezer after 72~h, after half of the water has turned into solid ice. This prevents cracks and bubbles from forming within the ice. The block is then cut using a chainsaw to the desired dimensions and its surface can be smoothed with a warm metal plate. The ice blocks used for the experiments on inclined plate were 3 cm thick, 30 cm long and 10 cm wide.

The blocks of ice are taken out of the freezer for two hours prior to the beginning of the experiments. Given the long duration of our experiments, there exists no temperature gradient within the ice, whose temperature remains close to \SI{0}{\celsius}. 
XPS caps were made from commercial Styrodur plates. PVC caps were machined from rigid PVC. Plaster and cement caps were molded from a mix of water and “Paris plaster” or Portland cement Cem I 52.5 respectively, dried at ambient temperature for one month and covered with a thin layer of epoxy resin (ResinPro), which prevents them from absorbing liquid water. Their density were respectively $1200\pm 20$~kg$\cdot$m$^{-3}$ and $1830\pm 40$~kg$\cdot$m$^{-3}$. Granite caps came from a core sample of soil from Archignat, Allier, France with a density $2800\pm 100$~kg$\cdot$m$^{-3}$.

The ice blocks were placed in a box opened on the top and the front, made of 10~cm thick XPS plates. The sample was illuminated from the sides using LEDs. Pictures were taken using a D5600 Nikon with a 200~mm lens placed 3,5~m away from the blocks. \REV{The relative humidity in the room was $60\pm 5\%$.} The room temperature was monitored using a thermocouple located at the top side of the box, it remains between 21 and \SI{22}{\celsius}. It had been verified by following small air bubbles in the ice that there was no measurable melting from the bottom of the ice block in contact with the XPS surface. For each picture, the instantaneous profile of the ice block was computed using a home-made contour detection procedure in Python: \REV{a first profile is extracted the software ImageJ~\cite{schneider2012nih} from which 100 points are interpolated. The displacement of the profile is then computed between two images $i$ and $i+1$: for each point of the profile on image $i$, the displacement of the profile is computed from the maximum of image $i+1$ on a line locally perpendicular to the profile on image $i$. This procedure is then iterated on every images.}

\begin{figure}[htbp!]
  \centering
  \includegraphics[width=8cm]{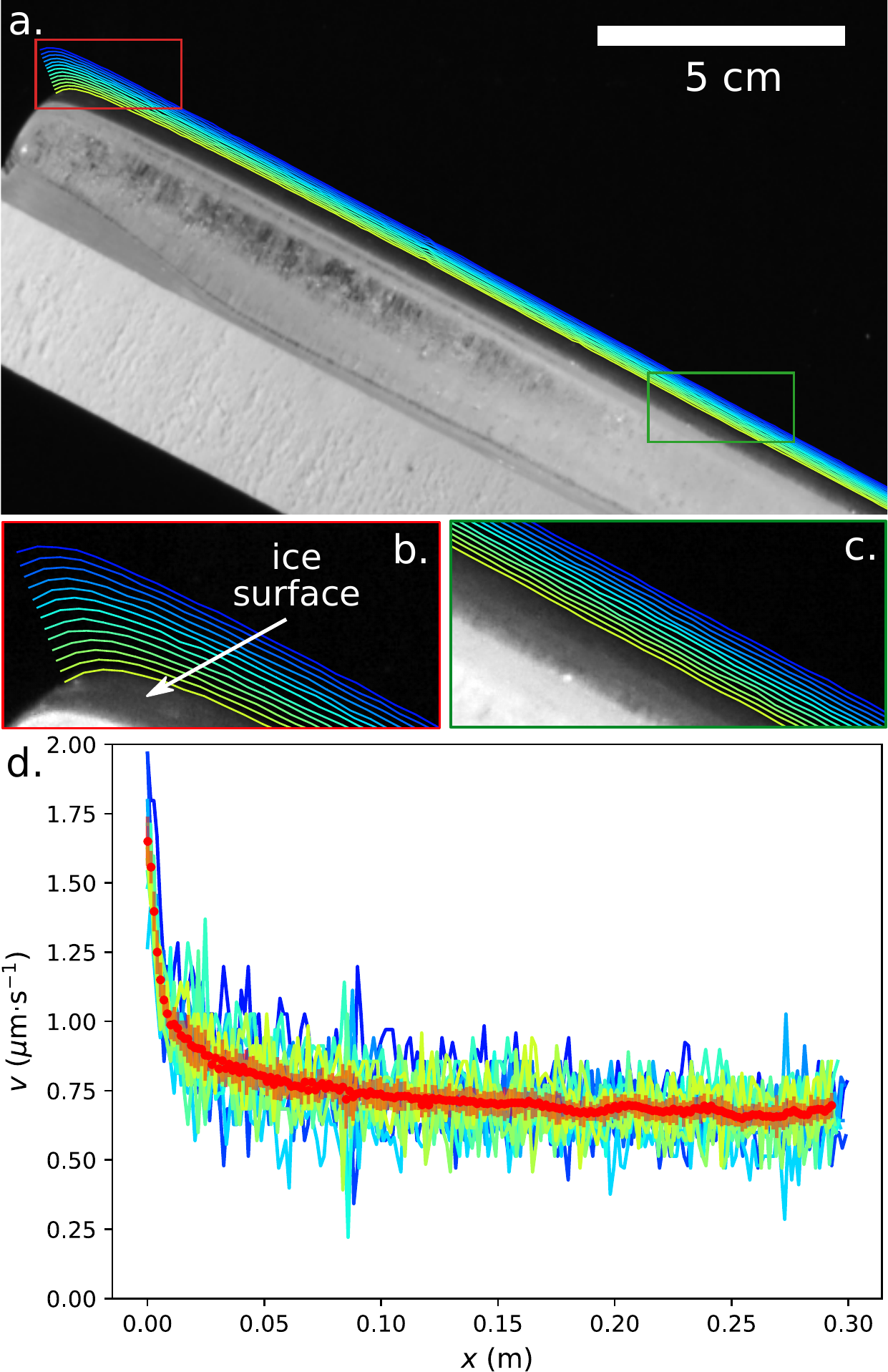}
 \caption{(a) Picture of an ice plate inclined with an angle $\theta=\ang{28}$ with respect to the horizontal. The colored contours correspond to the ice surface profile at regular time intervals (separated by $\mathrm{d} t = 8$~min). The red and green rectangles are magnified in (b) and (c). \REV{The bottom yellow profile corresponds to the displayed picture.} Velocity of the ice surface $v(x)=\mathrm{d}e(x)/\mathrm{d}t$ for each time step (solid lines) and averaged (red markers).}
  \label{fonte_incline_sup}
\end{figure}

 The local velocity of the ice $v_{\rm ice}$ is computed from the time evolution of the profile: the local displacement of the ice profile between two images is computed along the local normal vector with length $\mathrm{d}e(x)$, where $x$ is the arc length of the profile from the leading edge. The velocity of the ice surface is then computed as $v_{\rm ice}(x)=\mathrm{d}e(x)/\mathrm{d}t$. Figure~\ref{fonte_incline_sup} illustrates the measurement of the melting velocity profile of an inclined ice plate. 
\REV{Uncertainties were estimated statistically with a confidence interval of 90\%.}

%%%%%%%%%%%%%%%%%%%%%%%%%%%%%%%%%%%%%%%%
%%%%%%%%%%%%%%%%%%%%%%%%%%%%%%%%%%%%%%%%

\begin{figure}[htbp]
  \centering
  \includegraphics[width=8.6cm]{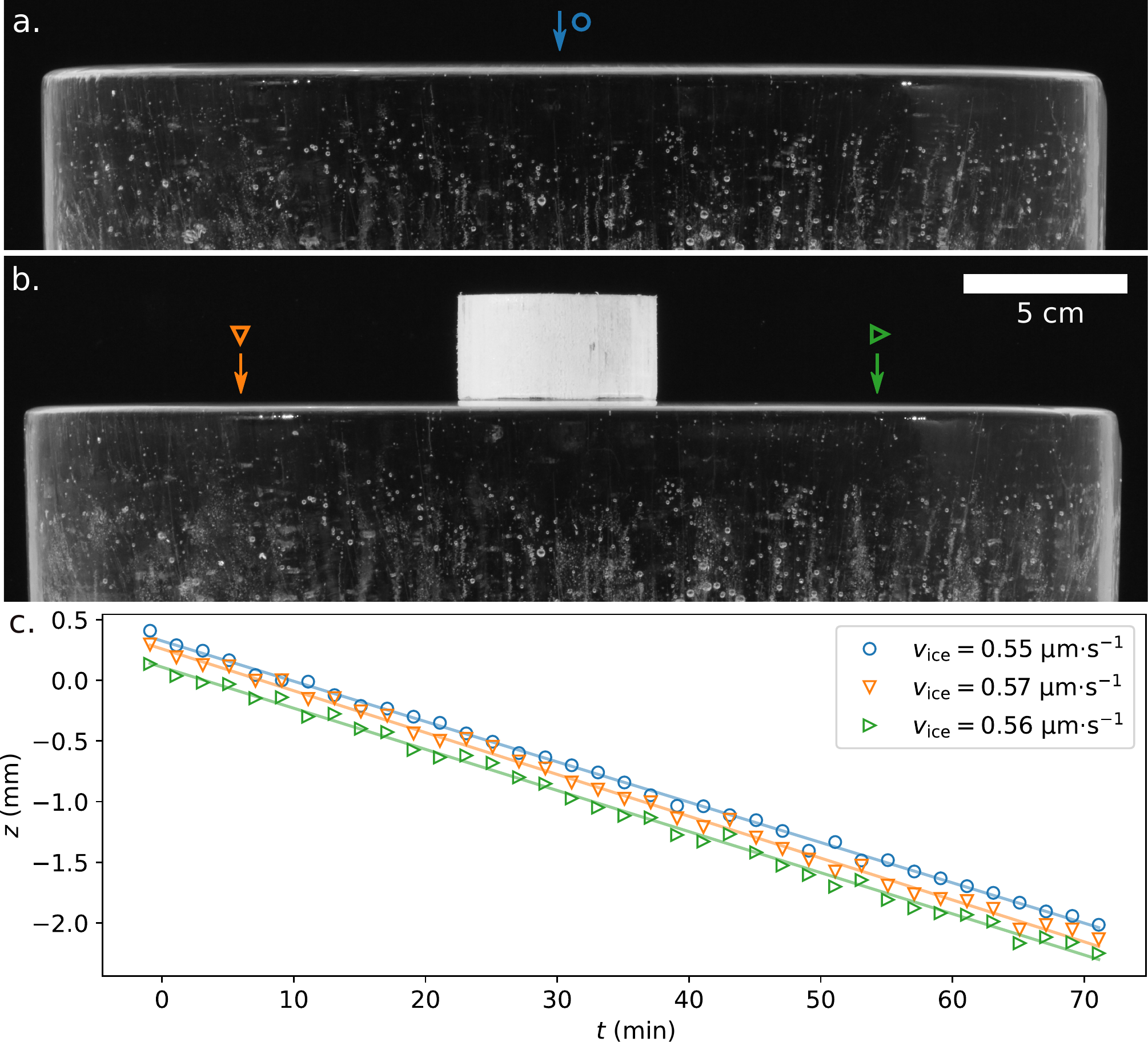}
 \caption{Measurement of the vertical air-ice interface velocity at the center of an ice block (a) and on the two sides of the same ice block on which a XPS cap is placed. (c) Evolution over time of the vertical position of the ice surface at the three points denoted with an arrow on picture a and b. The data were vertically shifted for visibility. The solid lines are linear fits from which a velocity $v_\mathrm{ice}$ is deduced.}
  \label{controle_v_glace}
\end{figure}

For each data point presented in Fig. 5 of the article, the air-ice interface velocity was measured approximately 5~cm away from the cap as shown in Fig.~\ref{controle_v_glace}.b. This velocity is identical to the one measured at the center of an ice block in the absence of a cap (see Fig.~\ref{controle_v_glace}.c).

%%%%%%%%%%%%%%%%%%%%%%%%%%%%%%%%%%%%%%%%
%%%%%%%%%%%%%%%%%%%%%%%%%%%%%%%%%%%%%%%%

\REV{\textbf{Convective heat flux on a inclined plate}}

The natural convection on the top surface of a cold inclined plate corresponds to a stable configuration, and is similar to the more commonly studied case of the air flow below a hot plate. %The expression of eq. 1 is valid in the laminar regime corresponding to Rayleigh numbers $\mathrm{Ra} \leq 10^9$, while in our experiment $\mathrm{Ra}_L \leq 2 \times 10^8$.
The prefactor $A$ in eq.~1 of the article can be computed analytically $A_\mathrm{th} = 0.508 \left(\dfrac{\mathrm{Pr}}{\mathrm{0.952+Pr}}\right)^{1/4} = 0.41$ where $\mathrm{Pr}=\dfrac{\nu_\mathrm{air}}{D_\mathrm{air}}=0.7$~\cite{lienhard_heat_2019}. It was also measured experimentally by Hassan and Mohamed~\cite{Hassan1970} in the case of a cold inclined plate facing upward ($A_\mathrm{exp} = 0.35$), by Fujii and Imura~\cite{FUJII1972} in the case of a hot inclined plate facing downward ($A_\mathrm{exp} = 0.56$) and by Churchill and Chu~\cite{CHURCHILL1975} who reported the following correlation  in the case of a vertical hot plate for $\mathrm{Ra}<10^9$ : $A_\mathrm{exp} = 0.670\left[1+\left(\dfrac{\mathrm{0.492}}{\mathrm{Pr}}\right)^{9/16}\right]^{-4/9}=0.51$.

\textbf{Thermal conductivities measurements}

\begin{figure}[htbp]
  \centering%
  \includegraphics[width=8.6cm]{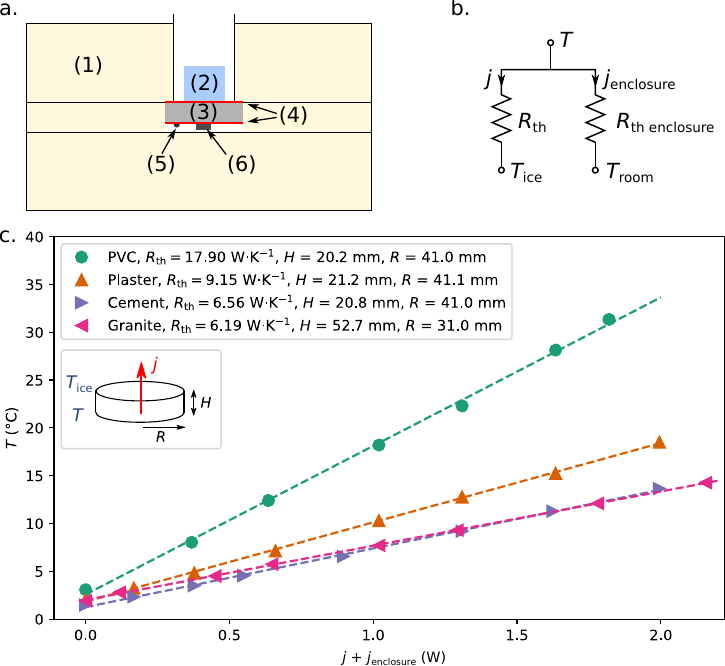}
 \caption{(a) Schematic of the experimental setup (1) XPS enclosure, (2) ice block, (3) cylindrical sample, (4) aluminum plates, (5) thermocouple, (6) heating resistor. (b) Electrical analogy of the heat transfer in the stationary regime. $R_\mathrm{th}$ and $R_\mathrm{th enclosure}$ correspond to the thermal resistance of the cylindrical sample and the XPS enclosure respectively. (c) Experimental results of the measured temperature $T$ as a function of the imposed heat transfer rate $j+j_\mathrm{enclosure}$ delivered by the heating resistor for cylindrical samples made of PVC, plaster and cement. \REV{The room temperature was $T_\mathrm{room} = 19.5$~\SI{}{\celsius} for PVC, plaster and cement and 24~\SI{}{\celsius} for granite.}} 
  \label{schema_mesure_conductivite}
\end{figure}

The thermal conductivities of PVC, plaster, cement and granite were measured using the experimental setup described in Fig.~\ref{schema_mesure_conductivite}.a. A cylindrical sample was sandwiched between two aluminum plates of negligible thermal resistance to insure an homogeneous temperature on both sides. The upper side was in contact with an ice block at $T_\mathrm{ice} = $~\SI{0}{\celsius}. The lower side, whose temperature $T$ was monitored using a thermocouple, was in contact with a heating resistor. A fraction $j$ of the thermal power generated by the resistor reaches the sample and a fraction $j_\mathrm{enclosure}$ is conducted through the surrounding XPS (see Fig.~\ref{schema_mesure_conductivite}.b). Figure~\ref{schema_mesure_conductivite}.c shows the experimental measurements of the temperature, $T$, as a function of the total heat transfer rate $j + j_\mathrm{enclosure}$. A linear fit $T = A (j + j_\mathrm{enclosure}) + B$ allows one to compute the thermal resistance of the sample $R_\mathrm{th} = \frac{A}{1-B/T_\mathrm{room}}$, with temperatures in \SI{}{\celsius}, and its thermal conductivity $\lambda = \frac{H}{R_\mathrm{th}\pi R^2}$. The results are summarized in Table~\ref{table_lambda} with the data gathered from the literature for XPS.

%Table~\ref{table_lambda} gather from the literature the values of the thermal conductivity of the materials used in this study.

%$$ T = A (j + j_\mathrm{enclosure}) + B $$

%$$ 1/A = \frac{1}{R_\mathrm{th}} + \frac{1}{R_\mathrm{th~enclosure}}$$

%$$ B/A = \frac{T_\mathrm{ice}}{R_\mathrm{th}}+\frac{T_\mathrm{room}}{R_\mathrm{th~enclosure}} $$

%%%%%%%%%%%%%%%%%%%%%%%%%%%%%%%%%%%%%%%%
%%%%%%%%%%%%%%%%%%%%%%%%%%%%%%%%%%%%%%%%

\begin{table}[h!]
\begin{center}
   \begin{tabular}{c| c| c |c| c}
Material & $\lambda$ [W$\cdot$m$^{-1}\cdot$K$^{-1}$] & ref. \\ \hline
XPS & $0.030 \pm 0.005$ & \cite{ctherm2020} \\
PVC & $0.21 \pm 0.01$ & measured \\ 
Plaster & $0.44 \pm 0.01$ & measured \\ 
Cement & $0.60 \pm 0.01$ & measured \\ 
Granite & \REV{$2.82 \pm 0.05$}  & \REV{measured} \\
   \end{tabular}
 \end{center}
\caption{Thermal conductivity of the materials constituting the caps used in this study.}
\label{table_lambda}
\end{table}

\end{document}